\documentclass[reprint,
amsmath,amssymb,
aps,
pre,
]{revtex4-2}
\usepackage{graphicx}
\usepackage{dcolumn}
\usepackage{bm}
\usepackage{float}
\usepackage{xcolor}
\usepackage{subfigure}
\usepackage[abs]{overpic}
\usepackage{hyperref}
\begin{document}
	\title{Crowding breaks the forward/backward symmetry of transition times in biased random walks}
	
	\author{Jaeoh Shin$^{1,2}$}
	
	\author{Alexander M. Berezhkovskii$^{5}$}
	
	\author{Anatoly B. Kolomeisky$^{1,2,3,4}$}
	\email{tolya@rice.edu}
	\affiliation{$^1$Department of Chemistry, Rice University, Houston, Texas, 77005, USA}
	\affiliation{$^2$Center for Theoretical Biological Physics, Rice University, Houston, Texas, 77005, USA}
	\affiliation{$^3$Department of Chemical and Biomolecular Engineering, Rice University, Houston, Texas, 77005, USA}
	\affiliation{$^4$Department of Physics and Astronomy, Rice University, Houston, Texas, 77005, USA}
	\affiliation{$^5$Mathematical and Statistical Computing Laboratory, Office of Intramural Research, Center for Information Technology, National Institutes of Health, Bethesda, Maryland 20892, USA}

	\begin{abstract}
		
		Microscopic mechanisms of natural processes are frequently understood  in terms of random walk models by analyzing local particle transitions. This is because these models properly account for dynamic processes at the molecular level and provide a clear physical picture. Recent theoretical studies made  a surprising discovery that in complex  systems the symmetry of molecular forward/backward transition times with respect to local bias in the dynamics may be broken and it may take longer to go downhill than uphill. The physical origins of these phenomena remain not fully understood. Here we explore in more detail the microscopic features of the symmetry breaking in the forward/backward transition times by analyzing exactly solvable discrete-state stochastic models. In particular, we consider a specific case of two random walkers on four-sites periodic lattice as the way to represent the general systems with multiple pathways. It is found that the asymmetry in transition times depends on several factors that include the degree of deviation from equilibrium, the particle crowding and methods of measurements of dynamic properties. Our theoretical analysis suggests that the asymmetry in transition times can be explored experimentally for determining the important microscopic features of natural processes by quantitative measuring the local deviations from equilibrium and the degrees of crowding.

	\end{abstract}
	
	\maketitle
	
	\section{Introduction}
	\label{intro}

	Numerous natural processes involve molecular motions in complex environments that reflect various inter-particle interactions \cite{bressloff2014stochastic,phillips2012physical,makarov2015single,chowdhury2000statistical}. Examples include ion transport through membrane channels \cite{cooper1985theory, roux2004theoretical},  motor proteins moving along cytoskeleton protein filaments \cite{kolomeisky2015motor}, protein-DNA interactions \cite{shvets2018mechanisms}. These phenomena are often modeled as random walks on lattices, providing important mechanical insights on the underlying microscopic processes \cite{,codling2008random,kolomeisky2015motor,shvets2018mechanisms,chowdhury2000statistical}. Generally, biased random walk models have been successfully employed for analyzing various processes in chemistry, physics and biology \cite{codling2008random,kolomeisky2015motor,chowdhury2000statistical,bressloff2014stochastic}. There are also significant recent advances in experimental studies of natural processes that allow researchers to monitor systems with high temporal and spatial resolutions \cite{yu2016single}.  The dynamics of the individual molecules can now be well monitored, for instance, by using fluorescent labeling methods \cite{shashkova2018systems,huhle2015camera,kolomeisky2015motor}. Among other important properties, transition times for molecular transitions from one state to another and drift velocities have been frequently measured with the goal to quantify the underlying microscopic processes. For example, for kinesin motor proteins that move along microtubules in {\it in vitro} single-molecule studies it was found that the transition times for forward and backward steps are the same, even though the forward transitions are predominant \cite{nishiyama2002chemomechanical}. This finding can be explained in terms of the microscopic reversibility of the elementary transition events for single particles, as shown in previous theoretical studies \cite{berezhkovskii2006identity,alvarez2006equivalence, makarov2015single}. 
	
	Interestingly, it was recently theoretically found that in some complex multi-particle systems the symmetry of uphill (against the local bias) and downhill (along the local bias) transition times can be broken. In contrast to expectations, it takes longer to go along the bias than against the bias. For example, such phenomena have been reported in the single-file diffusion \cite{ryabov2019counterintuitive}, in the biased random walk with exclusion interactions \cite{shin2020biased}, in the single-molecule motion with two different dynamic modes \cite{shin2020asymmetry} and in the particle dynamics on cycle processes with  strong coupling to the environment \cite{vorac2020cycle}. 
	It was  proposed that the asymmetry in the transition times is the sign that the system is out of equilibrium \cite{shin2020asymmetry}, as well as the consequence of the multi-particle interactions in the system \cite{ryabov2019counterintuitive,vorac2020cycle}. These theoretical studies, however, involved mostly numerical simulations or considered special simplified systems, raising the question if the observed phenomena are valid for more general cases.  In addition, the computer simulations were performed for systems with complex molecular interactions that made it difficult to clarify the underlying mechanisms. Thus, the microscopic picture behind the symmetry breaking for the single-particle dynamics in complex environments is still not fully understood.

	Consider a single particle that can move one site forward (backward) with a rate $u$ ($w$) on the otherwise empty one-dimensional lattice.  The probabilities for this biased random walker to step forward ($\Pi_{0}^{+}$) or backward ($\Pi_{0}^{-}$)  can be written as
	\begin{equation}
	\Pi_{0}^{+}=\frac{u}{u+w}, \quad \Pi_{0}^{-}=\frac{w}{u+w}.
	\end{equation}
	Its forward and backward mean transition times to the neighboring sites are the same and equal to its mean residence time on a site,
	\begin{equation}
	T_{0}^{+}=T_{0}^{-}=T_{0}=\frac{1}{u+w}.
	\label{Eq-2}
	\end{equation}
	The effective drift velocity for this particle is given by
	\begin{equation}
	V_{0}=\frac{\Pi_{0}^{+}-\Pi_{0}^{-}}{T_{0}}=(u-w).
	\end{equation}
	In real systems there are multiple particles that interact with each other, and this raises a question how specifically these single-particle features are affected by the presence of other particles in the system. In other words, how the molecular crowding modifies the measured properties of individual particles and what information about the system it provides.

	To answer these questions, in the present paper, we analyze the dynamics of a single tracer particle in a one-dimensional discrete-state stochastic model in the presence of other particles that interact with each other only via hard-core exclusion. This model considers multiple pathways for transitions, in contrast to our previous investigation \cite{shin2020biased}, and it allows us to explicitly quantify the dynamic properties of the single particle in a complex environment. Our results show that the asymmetry in the forward/backward transition time becomes more pronounced as the system deviates stronger from the equilibrium. Moreover, we found that the degree of the asymmetry depends on the method of measuring the transition times, which might be relevant for different experimental implementations. This is a new result that has not been reported before. In addition, it was found that crowding is an important factor that influences the asymmetry in local dynamic properties. This finding is in accordance with our recent work on simpler systems \cite{shin2020biased}. As detailed below, this current study generalizes the main results of the previous investigation by considering a more general transition network in which the dynamic properties are evaluated using two different approaches. Additionally, we argue that the degree of symmetry breaking in the transition times can be used to simultaneously measure the local degree of deviations from the equilibrium state of the system and the degree of crowding.

	\section{Theoretical Model}
	\label{Sec-model}
	
	To understand the microscopic features of the molecular motion in a complex environment, we utilize a random-walk approach for a one-dimensional discrete-state stochastic system, composed of a lattice of $L$ discrete sites where $N$ particles are moving. To simplify the analysis, we consider the periodic boundary conditions on the lattice. The particles can jump to the right and left with the rates $u$ and $w$, respectively, when the neighboring sites are not occupied.  While all the particles follow the same dynamics, to  analyze the molecular transitions we choose one of them as a tracer. This corresponds to the situation in experiments when the dynamics of one specific molecule, which might be fluorescently labeled or selectively chosen by optical tweezers, is followed \cite{shashkova2018systems,huhle2015camera}. We define a transition time for the tracer particle as the mean waiting time for the next allowed molecular transition. In the absence of other particles (no crowding), the transition time will be exponentially distributed with a single time scale that corresponds to the mean residence time in the given state. The situation is much more complex in the presence of multiple particles that affect each other dynamics.
	
\begin{figure}
		\centering
		(A)\includegraphics[width=0.85\columnwidth]{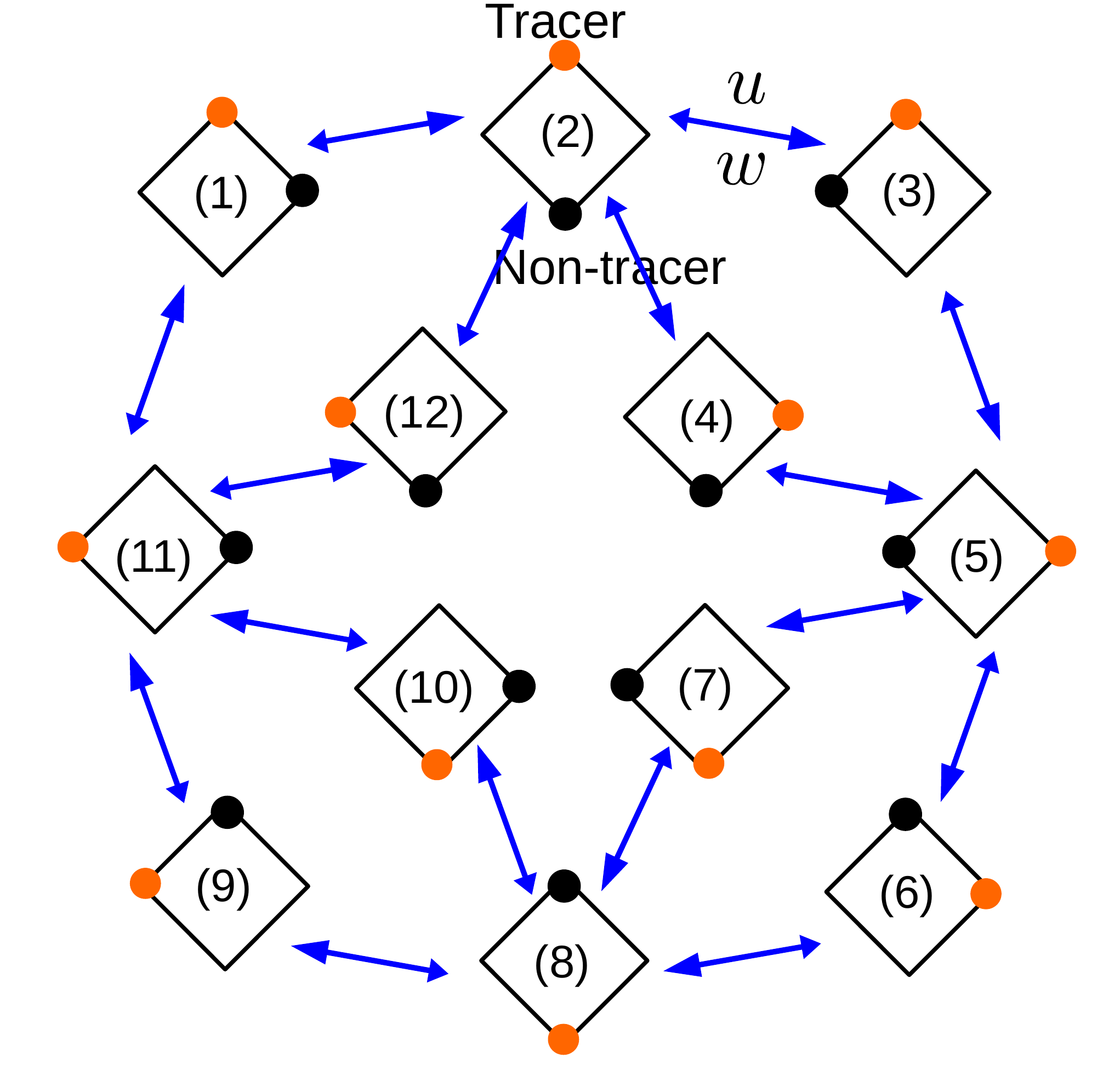}
		(B)\includegraphics[width=0.9\columnwidth]{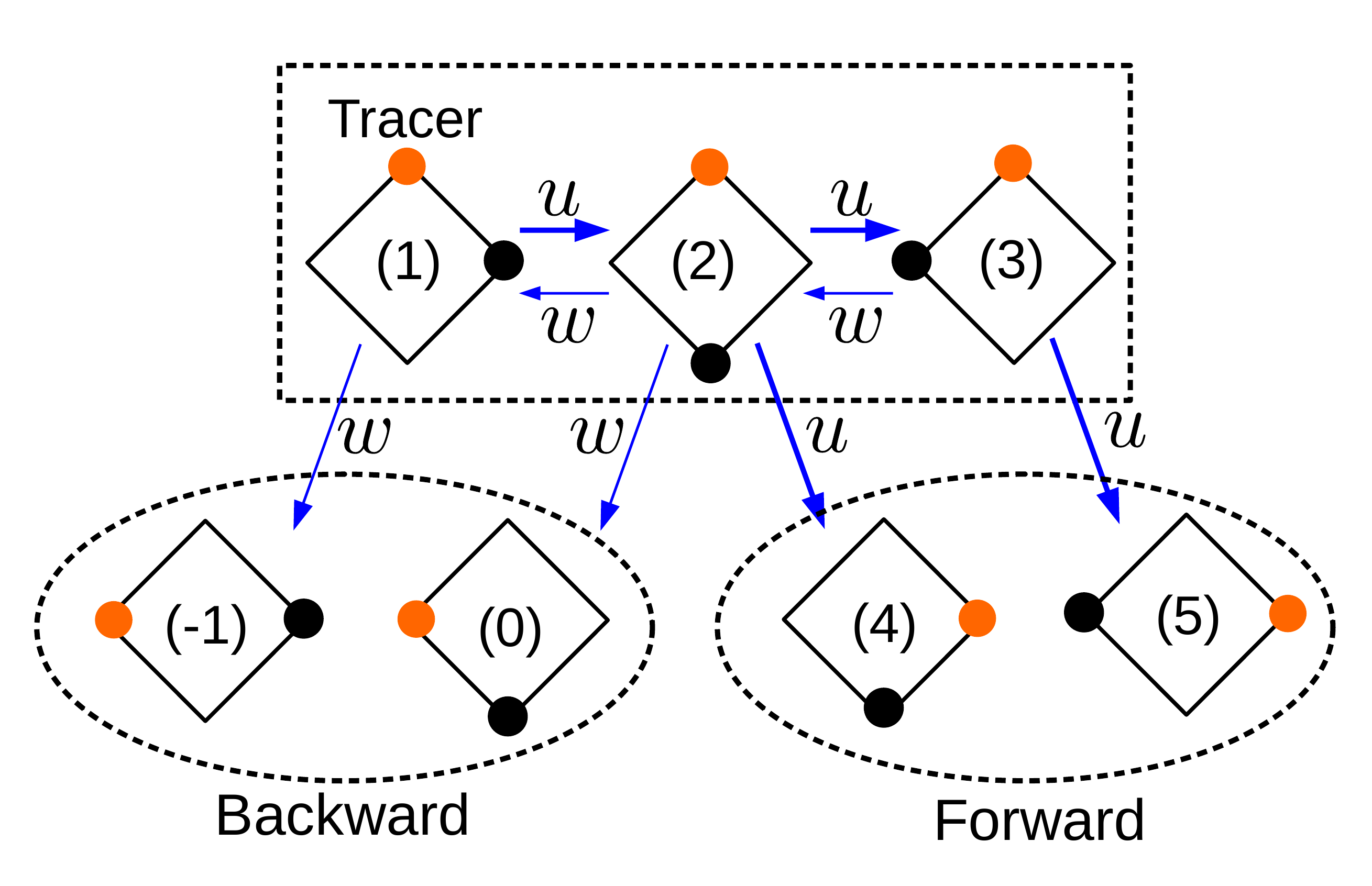}
		\caption{(A) A network of all possible discrete states in the system with two particles ($N=2$) on a lattice of $L=4$ sites with periodic boundary conditions. The transition rate along the bigger (smaller) head of arrows is $u$ ($w$). (B) A specific subset of transition pathways. Note that states (-1) and (0) here correspond to the states (11) and (12), respectively, in the panel A.}
		\label{Fig-1}
	\end{figure}

	In our recent work \cite{shin2020biased}, we considered a special case of the discrete-state model when the number of particles is $N=L-1$, i.e.,  there is only one empty site on the lattice. In this situation, the transition dynamics can be mapped onto a one-dimensional linear network of states that allows for a full quantitative analysis \cite{shin2020biased}. 
	For the general case of $L$ and $N<L-1$, however, the number of possible distinct particle configurations is larger, and the transition network is no longer one-dimensional. To understand the transition dynamics in such more complex systems, here we consider a special case with $L=4$ and $N=2$. 
	In this case, there are total 12 states as shown in Fig. \ref{Fig-1}A. It has a non-linear transition network structure and better represents a more general situation. As the system has four-fold symmetry, we can describe  a single transition event by considering a sub-system shown in Fig. \ref{Fig-1}B. For a given tracer position, there are three different configurations of the system as labeled by the states $(1)$, $(2)$, and $(3)$. The tracer can make forward (backward) transition by stepping into the configuration (4) or (5) [(-1) or (0)]: see Fig. \ref{Fig-1}B. Therefore, to calculate the mean transition time, we need to consider a total of six possible paths for both forward and backward transitions (three possible initial configurations and two possible final configurations).


	Let us consider first the forward transition path via the stepping to the state (5). We define the first-passage probability density function $F_{i,5}(t)$ for the initial state $i$ [$i=(1)$, $(2)$, or $(3)$] as a probability to reach the state $(5)$ at time $t$ for the first time before reaching any of the states $(-1),(0),(4)$ starting at $t=0$ at the state $i$. This function is the probability flux entering the state (5) at time $t$. The dynamics of these functions are governed by the backward master equations \cite{shin2020biased},
	\begin{equation}
	\frac{d F_{1,5}(t)}{dt}=-(u+w)F_{1,5}(t)+uF_{2,5}(t),
	\end{equation}
	\begin{equation}
	\frac{d F_{2,5}(t)}{dt}=-2(u+w)F_{2,5}(t)+wF_{1,5}(t)+uF_{3,5}(t),
	\end{equation}
	\begin{equation}
	\frac{d F_{3,5}(t)}{dt}=-(u+w)F_{3,5}(t)+wF_{2,5}(t)+uF_{5,5}(t).
	\end{equation}
	In addition, we have $F_{5,5}(t)=\delta(t)$, which means that if the initial state is already in the state $(5)$, the process will end immediately.
	
	To solve these equations, we apply the Laplace transformation, $\widetilde{F}_{i}(s)\equiv \int_{0}^{\infty} F_{i}(t)\exp(-st)dt$, where $s$ is the Laplace variable. Then the above differential equations modify into simpler algebraic equations,
	\begin{equation}
	(s+u+w)\widetilde{F}_{1,5}(s)=u \widetilde{F}_{2,5}(s),
	\end{equation}
	\begin{equation}
	(s+2u+2w)\widetilde{F}_{2,5}(s)=w \widetilde{F}_{1,5}(s)+u \widetilde{F}_{3,5}(s),
	\end{equation}
	\begin{equation}
	(s+u+w)\widetilde{F}_{3,5}(s)=w \widetilde{F}_{2,5}(s)+u.
	\end{equation}

	These algebraic equations can be easily solved, allowing us to obtain the full description of the transition dynamics, i.e., all functions $\widetilde{F}_{i,5}(s)$, for $i=1$, $2$, $3$. Our main focus here is on the splitting probabilities $\Pi_{i,5}$, which are the probabilities for the tracer particle starting from the state $i$ to end up in the state $5$, and the mean transition times $T_{i,5}$ for such events. In order to calculate those quantities, we expand the Laplace transform of $F_{i,5}(t)$ in small $s$, $\widetilde{F}_{i,5}(s)=\int_{0}^{\infty}[1-st]F_{i,5}(t)dt+\mathcal{O}(s^2)$. 
	The splitting probability can be obtained using the relation $\Pi_{i,5}\equiv\int_{0}^{\infty}F_{i,5}(t)dt= \lim_{s\rightarrow 0}\widetilde{F}_{i,5}(s)$. The result is
	\begin{equation}
	\Pi_{1,5}=\frac{u^3}{2(u+w)(u^2+uw+w^2)};
	\end{equation}
	\begin{equation}
	\Pi_{2,5}=\frac{u^2}{2(u^2+uw+w^2)};
	\end{equation}
	\begin{equation}
	\Pi_{3,5}=\frac{u(2u^2+3uw+2w^2)}{2(u+w)(u^2+uw+w^2)}.
	\end{equation}
	
	To find the mean transition time $T_{i,5}$, we notice that this is a conditional mean first-passage time. Then we utilize the fact that the probability distribution function of this time $\phi_{i,5}(t)$ is the ratio of the function $F_{i,5}(t)$ to the splitting probability $\Pi_{i,5}$, $\phi_{i,5}(t)=F_{i,5}(t)/\Pi_{i,5}$. 
	This allows us to calculate the mean transition times $T_{i,5}$ using the equation
	\begin{equation}
	T_{i,5}\equiv\int_{0}^{\infty} t \phi_{i,5}(t) dt = \frac{1}{\Pi_{i,5}}\int_{0}^{\infty}t F_{i,5}(t)dt= \frac{1}{ \Pi_{i,5}}\frac{-\partial \widetilde{F}_{i,5}(s)}{\partial s} \vert_{s \rightarrow 0}.
	\end{equation}
	The results are,
	\begin{equation}
	T_{1,5}=\frac{5u^2+8uw+5w^2}{2(u+w)(u^2+uw+w^2)}
	\end{equation}
	\begin{equation}
	T_{2,5}=\frac{3(u+w)}{2(u^2+uw+w^2)}
	\end{equation}
	\begin{equation}
	T_{3,5}=\frac{4u^4+13u^3w+20u^2w^2+13uw^3+4w^4}{2(u+w)(u^2+uw+w^2)(2u^2+3uw+2w^2)}.
	\end{equation}
	Note that these transition times are symmetric with respect to the change of $u$ and $w$, which means that the mean time for the transition from the {\it particular} initial state to the {\it particular} final state is the same as the mean time for the reversed path \cite{shin2020biased}.

	Similarly, we consider the forward transitions to the state (4). In this case, the state (4) is now an absorbing boundary, and the initial condition is $F_{4,4}=\delta(t)$. Using the same formalism, we obtain the splitting probabilities and the mean transition times,
	\begin{equation}
	\Pi_{1,4}=\frac{u^2}{2(u^2+uw+w^2)};
	\end{equation}
	\begin{equation}
	\Pi_{2,4}=\frac{u(u+w)}{2(u^2+uw+w^2)};
	\end{equation}
	\begin{equation}
	\Pi_{3,4}=\frac{uw}{2(u^2+uw+w^2)};
	\end{equation}
	and
	\begin{equation}
	T_{1,4}=T_{3,4}=\frac{3(u+w)}{2(u^2+uw+w^2)};
	\end{equation}
	\begin{equation}
	T_{2,4}=\frac{u^2+4uw+w^2}{2(u+w)(u^2+uw+w^2)};
	\end{equation}
	Following the symmetry arguments, corresponding results for the backward transitions to the states $(-1)$ and $(0)$ can be obtained by exchanging $u$ and $w$.
	
	The experimental measurements of transition dynamics report the average over all possible pathways.  This can be fully accounted in our theoretical approach. We define weight factors $p_{i,\alpha}$ as the probability for the system to start in the state $i$ at time $t=0$, and the subscript $\alpha$ labels the method of measuring of dynamic properties. Then the average forward and backward probabilities of stepping for the tracer particle are given as
	\begin{equation}
	\Pi_{\alpha}^{+}=\sum_{i=1}^{3} p_{i,\alpha} (\Pi_{i,4}+\Pi_{i,5}), \quad \Pi_{\alpha}^{-}=\sum_{i=1}^{3} p_{i,\alpha} (\Pi_{i,-1}+\Pi_{i,0}).
	\end{equation}
	Similarly, the mean forward (backward) transition times are obtained by averaging over all possible forward (backward) transition paths, leading to
	\begin{equation}\label{eq24}
	T_{\alpha}^{+}=\frac{1}{\Pi_{\alpha}^{+}} \sum_{i=1}^{3} p_{i,\alpha}(\Pi_{i,4} T_{i,4} +\Pi_{i,5} T_{i,5}),
	\end{equation}
	and
	\begin{equation}
	T_{\alpha}^{-}=\frac{1}{\Pi_{\alpha}^{-}} \sum_{i=1}^{3} p_{i,\alpha}(\Pi_{i,-1} T_{i,-1} +\Pi_{i,0} T_{i,0}).
	\end{equation}
	These expressions can be understood in the following way. The quantities $p_{i,\alpha}\Pi_{i,4}/\Pi_{\alpha}^{+}$ and $p_{i,\alpha}\Pi_{i,5}/\Pi_{\alpha}^{+}$ can be viewed as weights for the forward pathways that start at the state $i$  and end in the state (4) and (5), respectively. Similarly, the quantities  $p_{i,\alpha}\Pi_{i,-1}/\Pi_{\alpha}^{-}$ and $p_{i,\alpha}\Pi_{i,0}/\Pi_{\alpha}^{-}$ can be viewed as weights for the backward pathways that start in the state $i$ and end in the state (-1) and (0), respectively.
	
	Finally, we can also evaluate the average residence time at each lattice site,
	\begin{equation}
	T_{\alpha}^{res}=\Pi_{\alpha}^{+} T_{\alpha}^{-}+\Pi_{\alpha}^{+} T_{\alpha}^{-},
	\end{equation}
	and the average drift velocity of the tracer particle,
	\begin{equation}\label{eq27}
	V_{\alpha}=\frac{\Pi_{\alpha}^{+}-\Pi_{\alpha}^{-}}{T_{\alpha}^{res}}.
	\end{equation}
	
	In the following, we set the forward hopping rate $u=1$ s$^{-1}$, and only the backward hopping rate $w$ ($\le u$) is varied. It is also convenient to introduce a bias parameter $x\equiv w/u$ that quantifies the tendency of the random walkers to move in different directions. For $x=1$ we have an unbiased random walk, and the system is in equilibrium, while $0 \le x<1$ corresponds to biased random walks under non-equilibrium conditions. The times below are measured  in the units of $u^{-1}$. Then all the quantities of interest can be expressed in terms of $x$, for instance, $\Pi_{0}^{+}=\frac{1}{1+x}$, $\Pi_{0}^{-}=\frac{x}{1+x}$, $V_{0}=1-x$, and $T_{0}=\frac{1}{1+x}$.

	\section{Results}
	\label{Sec-results}

	Since in our approach the tracer particle dynamics can be fully quantified for any pathway in the system, the effect of crowding on the symmetry breaking of transition times can now be explicitly evaluated. However, we notice that in experiments there could be several different ways of measuring these dynamic properties. Two different approaches are identified here, although other methods of measurements might also be possible. 
	
	In the first case, which we label as a resetting average approach, when the tracer particle makes a transition the system is immediately reset to the random initial state. This means that the system can start with equal probability in any state $i$, $p_{i,r}=\frac{1}{3}$ ($\alpha=r$ corresponds to the resetting method). Using Eqs. (10)-(12), (17)-(19) and (23), we obtain the expressions for the forward and backward stepping probabilities,
	\begin{equation}
	\Pi_{r}^{+}=\Pi_{0}^{+} \frac{3+4x+2x^{2}}{3(1+x+x^{2})}=\Pi_{0}^{+} \left[1+\frac{x(1-x)}{3(1+x+x^{2})}\right],
	\end{equation}
	and 
	\begin{equation}
	\Pi_{r}^{-}=\Pi_{0}^{-} \frac{2+4x+3x^{2}}{3(1+x+x^{2})}=\Pi_{0}^{-} \left[1-\frac{(1-x)}{3(1+x+x^{2})}\right],
	\end{equation}
	where $x=w/u$ and  $\Pi_{0}^{\pm}$ are the corresponding probabilities when there is no crowding [see Eq.(\ref{Eq-1})]. We immediately notice  that  here crowding affects the stepping probabilities because  $\Pi_{r}^{+}>\Pi_{0}^{+}$ and $\Pi_{r}^{-}<\Pi_{0}^{-}$. The probability to step in the forward (backward) direction increases (decreases) in comparison with the no-crowding situation. 
	
	Similar calculations for the forward and backward transition times lead to
	\begin{equation}
	T^{+}_{r}=T_{0}\frac{8 + 24 x + 31 x^{2} + 17 x^3 + 4 x^4 }{2 (1+x + x^{2}) (3+4x+2x^{2})},
	\end{equation}
	and 
	\begin{equation}
	T^{-}_{r}=T_{0}\frac{4 + 17 x + 31 x^{2} + 24 x^3 + 8 x^4 }{2 (1+x + x^{2}) (2+4x+3x^{2})},
	\end{equation}
	where $T_{0}=1/(1+x)$ is the mean transition time in the case of no crowding [see Eq. (\ref{Eq-2})]. Here we have $T_{r}^{+}>T_{r}^{-}>T_{0}$, which means that crowding affects both forward and backward mean transition times by making them longer in comparison with the no-crowding situation. Note that our calculations also lead to an unexpected result: the transition against the bias occurs faster than the transition along the bias. The mean residence time at any site for the tracer particle can now be evaluated as
	\begin{equation}
	T_{r}^{res}=\Pi_{r}^{+} T_{r}^{+} + \Pi_{r}^{-} T_{r}^{-}= T_{0} \frac{2(2+3x+2x^{2})}{3(1+x+x^{2})},
	\end{equation}
	suggesting that crowding increases the residence times. 
	
	One could also estimate the drift velocity for the tracer particle in the resetting method using Eq. (\ref{eq27})
	\begin{equation}
	V_{r}=\frac{\Pi_{r}^{+}-\Pi_{r}^{-}}{T_{r}^{res}}=V_{0} f_{r}(x),
	\end{equation}
	where $V_{0}$ is the drift velocity in the no-crowding case. The function
	\begin{equation}
	f_{r}(x)=\frac{3+5x+3x^{2}}{2(2+3x+2x^{2})}
	\end{equation}
	monotonically increases from $0.75$ at $x=0$ to $\frac{11}{14}$ at $x=1$. One can see that crowding, as expected, lowers the effective drift velocity ($V_{r}<V_{0}$).

\begin{figure}
		\centering
		(A)\includegraphics[width=0.95\columnwidth]{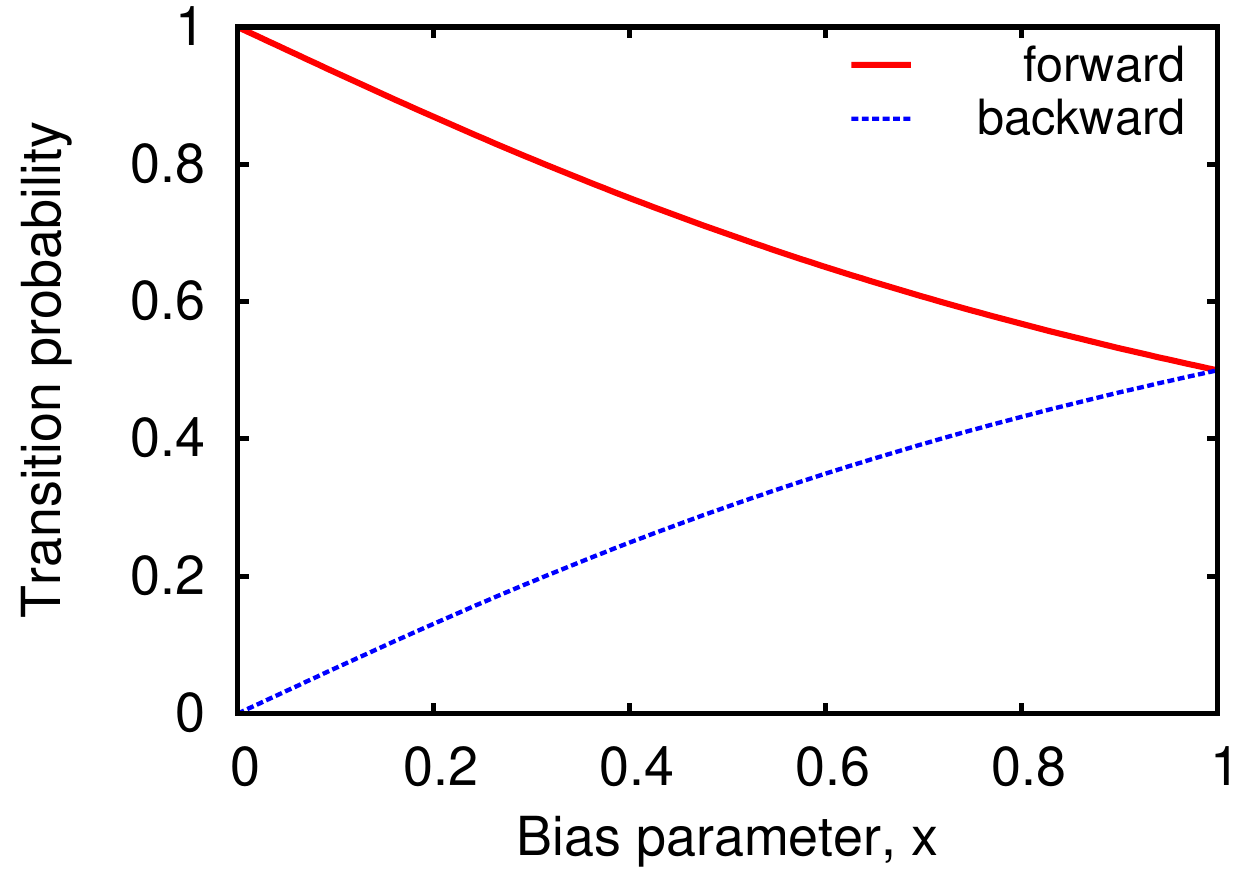}
		(B)\includegraphics[width=0.95\columnwidth]{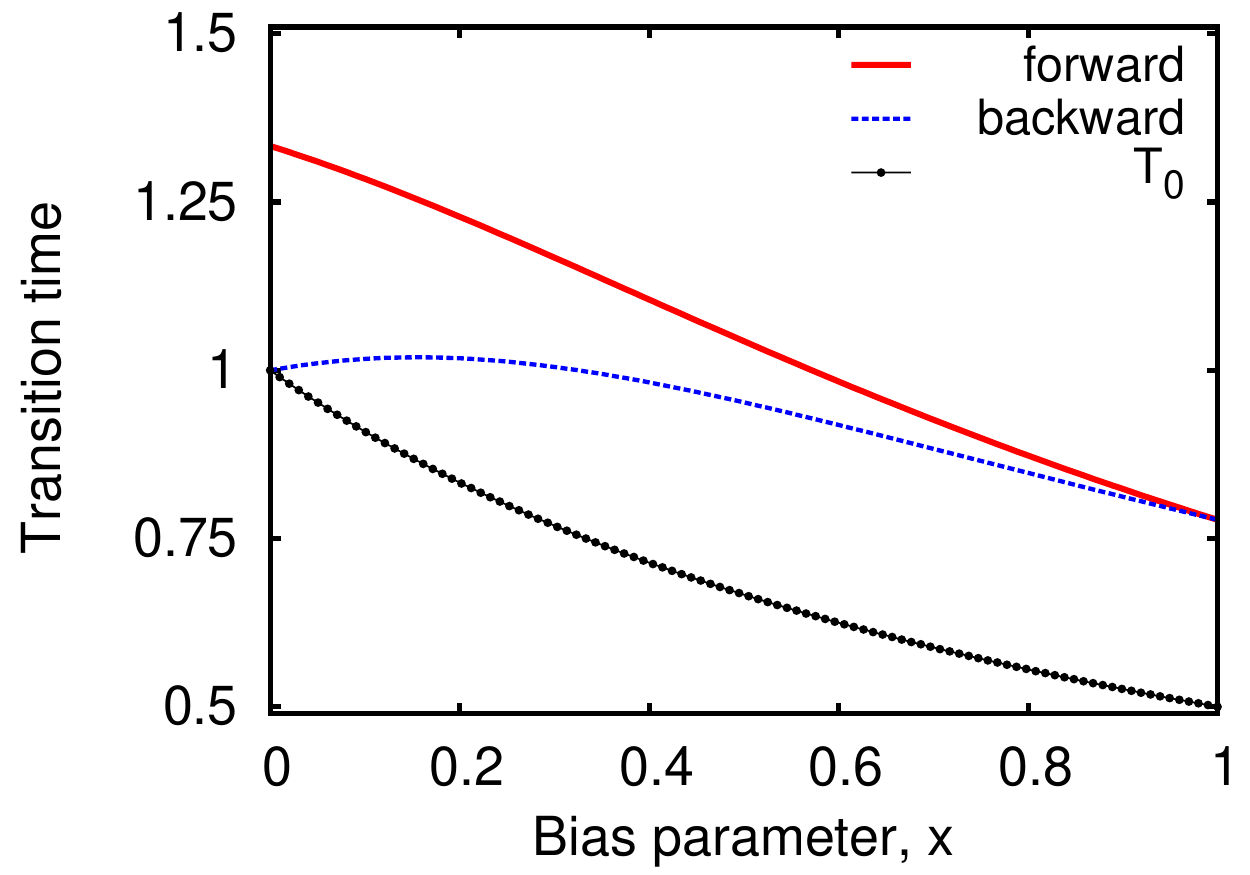}
		\caption{Transition probabilities and transition times for the model with $L=4$ and $N=2$ for resetting average case. (A) Transition probability (B) Transition time. Red solid lines are for the forward transition probability and the mean transition time, and blue dashed lines are for the backward transition probability and the mean transition time. Black dotted lines show the mean residence time $T_{0}$.}
		\label{Fig-2}
	\end{figure}

	In Fig. \ref{Fig-2} we show the dynamic properties of the tracer particle calculated using the resetting  averaging approach for different degrees of bias. As expected, for $x=1$ the system is in equilibrium, leading to the same stepping probabilities ($\Pi_{r}^{+}=\Pi_{r}^{-}=0.5$) and equal mean transition times in both directions  ($T_{r}^{+}=T_{r}^{-}$), although both mean transition times are still longer than  the mean transition times without crowding ($T_{0}$). However, when the system deviates from equilibrium ($0 \le x <1$) there are unexpected changes in the dynamic properties. While the probability for the forward transition increases, this does not lead to naively expected lowering of the mean forward transition times. Instead, $T_{r}^{+}$ increases and, surprisingly, the mean backward transitions are now always faster ($T_{r}^{+} > T_{r}^{-}$). This is clearly the effect produced by crowding in the system.

	In the second scenario of possible measurements of dynamic properties, which we call a time (or trajectory) averaging approach (now the subscript $\alpha=t$),  the system might be continuously observed for a very  long time \cite{shin2020biased}. Then all successful events, the forward or the backward steps of the tracer particle, are counted and the mean transition times in each direction are obtained. In this case, the weight factors $p_{i,t}$ are determined via the relations,
	\begin{equation}\label{eq22}
	p_{1,t}=p_{1,t}\Pi_{1,4}+p_{2,t}\Pi_{2,4}+p_{3,t}\Pi_{3,4};
	\end{equation}
	\begin{eqnarray}\label{eq23}
	p_{2,t}= & & \nonumber \\
	p_{1,t}(\Pi_{1,-1}+\Pi_{1,5})+p_{2,t}(\Pi_{2,-1}+\Pi_{2,5}) & & \nonumber \\
	+p_{3,t}(\Pi_{3,5}+\Pi_{3,-1}); & &
	\end{eqnarray}
	\begin{equation}\label{eq24}
	p_{3,t}=p_{1,t}\Pi_{1,0}+p_{2,t}\Pi_{2,0} +p_{3,t}\Pi_{3,0}.
	\end{equation}
	These expressions can be explained  using the following arguments. The state (1) corresponds to the situation when the tracer particle is behind the crowding particle in the clockwise direction (see Fig. \ref{Fig-1}A). To start in this configuration, in the previous transition the system must end up in the state (4) [Note that the state (1) and the state (4) can be considered to be identical since they can be transformed into each other via rotation]. This can only happen via transitions $1 \rightarrow 4$, $ 2 \rightarrow 4$ and $3 \rightarrow 4$ with the corresponding weights. This leads to Eq. (\ref{eq22}). The system will start in the state (2) if the previous transitions were $1 \rightarrow -1$, $1 \rightarrow 5$, $2 \rightarrow -1$, $2 \rightarrow 5$, $3 \rightarrow -1$ or $3 \rightarrow 5$ (Fig. \ref{Fig-1}A). This explains Eq. (\ref{eq23}). Finally, to start in the state (3) the previous transitions must end up in the state (0), which leads to the Eq. (\ref{eq24}).

	Solving Eqs. (\ref{eq22}), (\ref{eq23}) and (\ref{eq24}), we obtain 
	\begin{equation}
	p_{1,t}=\frac{1}{2(1+x)}, \quad p_{2,t}=\frac{1}{2}, \quad p_{3,t}=\frac{x}{2(1+x)},
	\label{eq25}
	\end{equation}
	which allow us to evaluate the dynamic properties of the tracer particle. Specifically, we obtain
	\begin{equation}
	\Pi_{t}^{+}=\Pi_{0}^{+}=\frac{1}{1+x}, \quad \Pi_{t}^{-}=\Pi_{0}^{-}=\frac{x}{1+x}.
	\end{equation}
	This result suggest that, in contrast to the resetting method, in the time-averaging approach the crowding does not influence the probabilities to step forward or backward. However, this is not the case for transition times for which we obtain
	\begin{equation}
	T_{t}^{+}=T_{0} \frac{3+4x+2x^{2}}{2(1+x+x^{2})}, \quad T_{t}^{-}=T_{0} \frac{2+4x+3x^{2}}{2(1+x+x^{2})}.
	\end{equation}
	One can see that crowding slows down transition dynamics $T_{t}^{+}> T_{t}^{-}>T_{0}$. Again, the transitions against the bias occur faster than those along the bias. We can also easily evaluate the mean residence time
	\begin{equation}
	T_{t}^{res}=\frac{3}{2} T_{0},
	\end{equation}
	and the drift velocity
	\begin{equation}
	V_{t}=\frac{2}{3} V_{0}.
	\end{equation}
	As expected, in the case of crowding the residence times are longer while the drift velocity is slower. It is interesting to note, however, that although the trends are the same, the two different methods of measurements (averaging) lead to slightly different estimates of the dynamic properties.

\begin{figure}
		\centering
		(A)\includegraphics[width=0.95\columnwidth]{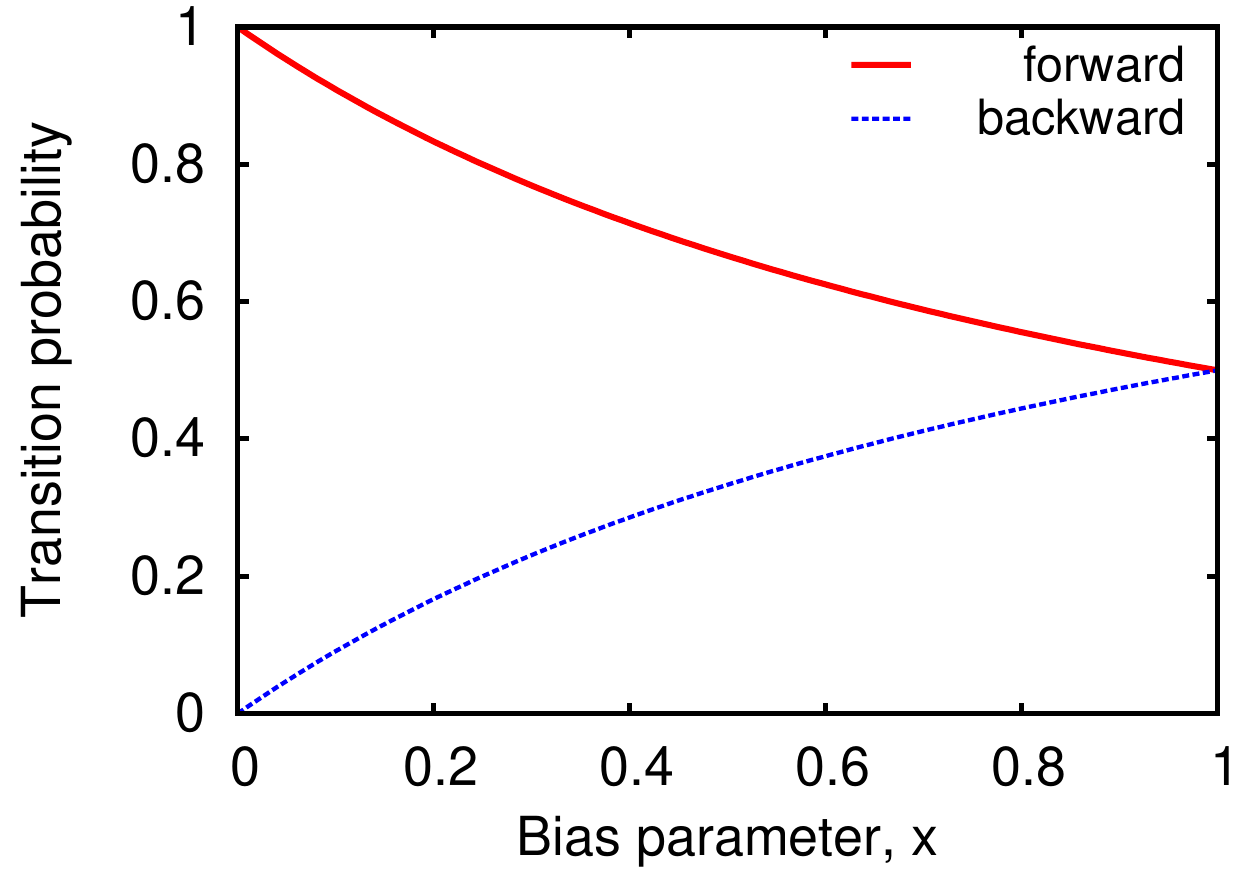}
		(B)\includegraphics[width=0.95\columnwidth]{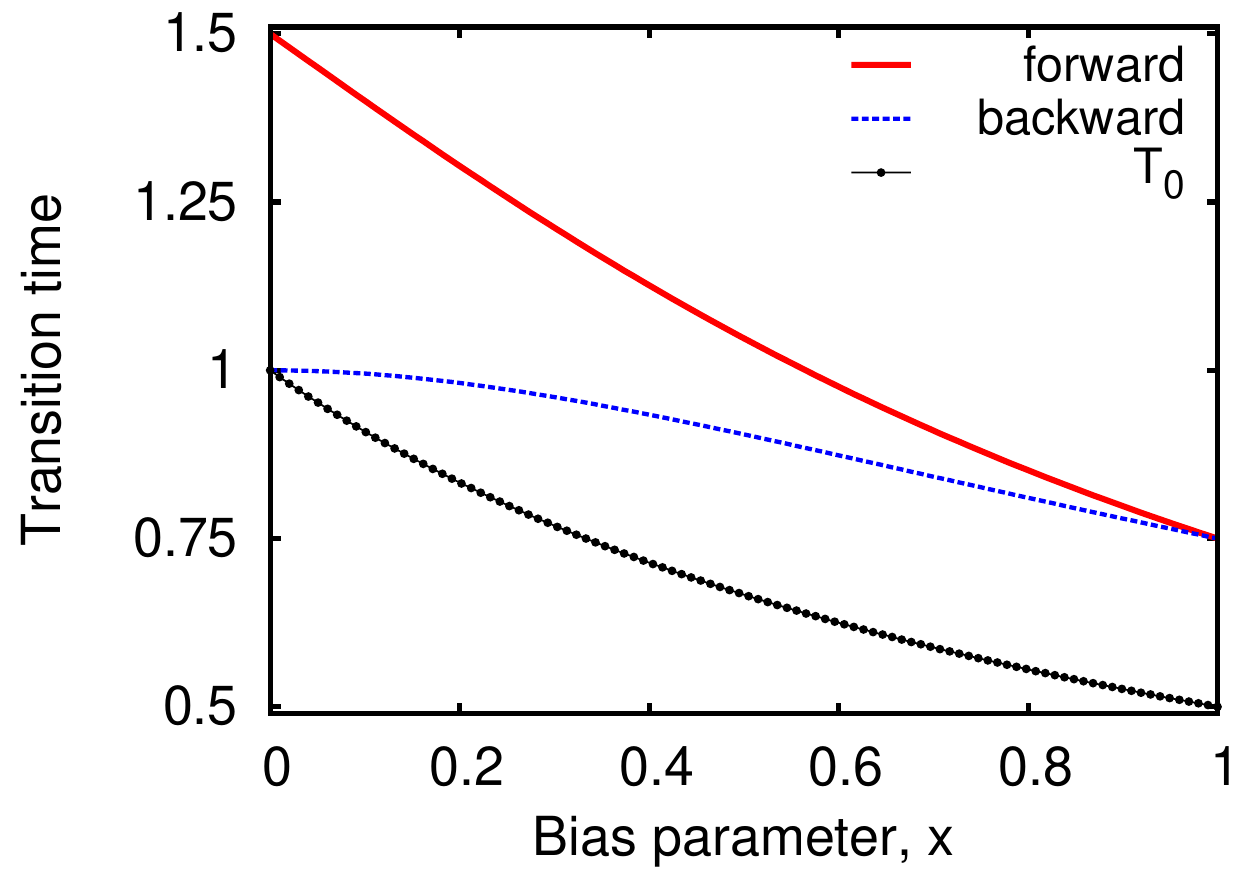}
		\caption{Transition probabilities and transition times for the model with $L=4$ and $N=2$ for time average case. (A) Transition probability (B) Transition time. Red solid lines are for the forward transition probability and the mean transition time, and blue dashed lines are for the backward transition probability and the mean transition time. Black dotted lines show the mean residence time $T_{0}$.}
		\label{Fig-3}
	\end{figure}

	Fig. \ref{Fig-3} shows the dynamic properties of the tracer particle calculated using the time-average approach. One can see that for the equilibrium situation ($x=1$) both stepping probabilities are the same and both mean transition times are equal to each other. Deviations from the equilibrium ($x<1$) increase the probability of the forward transitions, but again it does not lead to the acceleration. The mean forward transition times increase with the degree of the bias (lower $x$), and they are always longer than the backward transition times.

	It can be clearly seen from Figs. 2 and 3 that for both methods of measuring the dynamic properties the transition times generally are not the same. The mean backward transition times for both methods agree only for the special situations with $x \rightarrow 0^{+}$. The non-monotonic behavior of the mean backward transition time for the resetting method is observed (Fig 2B), while for the time-averaging method this time decreases monotonically with $x$ (Fig 3B). Larger deviations are found for the forward transition times calculated by different methods, especially for $x \rightarrow 0$ where $T^{+}_{r}/T^{+}_{t} \sim 8/9$.

	The differences between ``measured" transition dynamics for the tracer particle in two approaches can be explained using the following arguments. In the resetting method, each successful event corresponds to only one transition. This means that all initial states are probed with equal probability. This is not the case, however, for the time-averaging method. During the long-time measurements, the initial states more frequently are those that correspond to more probable particle configurations. One can also explain this effect using the Fig. \ref{Fig-1}A. For the bias in the clockwise motion ($u>w$, $x<1$), the successful transitions of the tracer particle end up more in the state (4) and (5) than in the state (-1) and (0), and this corresponds to the initial states to be more frequently the state (1) or (2), but not the state (3).
	
	The two different measuring methods are related to  ``time-averaged" versus ``ensemble-averaged" approaches to estimate the relevant properties in statistical mechanics. In Ref. \cite{hou2018biased}, the authors investigated the intriguing differences between these two types of measurements and ergodic properties by considering the mean-squared displacements of the biased continuous-time random walks. It will be interesting to extend it to the systems considered in our work. Note, however, that there is no ergodicity breaking in our system of two random walkers on the lattice with four sites. 
	
	Our calculations indicate that the asymmetry of transition times, defined by the ratio of forward and backward transition times, $A_{\alpha}\equiv\frac{T_{\alpha}^{+}}{T_{\alpha}^{-}}$, is larger for the time-averaged approach ($A_{t}>A_{r}$): see Figs. 2 and 3. The reason is that in this case the probability of being in the state (1) at time $t$=0 ($p_{1,t}$), from which the forward transition time is the longest, is larger than $p_{3,t}$, from which the forward transition time is the shortest. This is because the forward transitions are more frequent than the backward transition, leading to the preference for the initial state to be (1) or (2). But for the resetting method all initial states are equally probable.

	Although both methods of measuring/averaging give slightly different results for the transition times and probabilities, the general trends are mostly the same. It is found that  increasing the bias (lowering $x$) leads to higher forward transition probabilities and smaller backward transition probabilities (Figs. 2A and 3A). At the same time, the mean transition time along the bias (in the forward direction) is always longer than that in the backward direction: see Figs. 2B and 3B. This result is unexpected since one would expect a faster motion in the bias direction, especially if the forward probabilities are larger, but this does not happen.
	
	To better understand these observations, let us consider a limiting case of $u \gg w$ ($x \ll 1$). In this situation, the tracer particle can make a forward transition from all the three initial states with a finite probability, $\mathcal{O}(1)$ (see Fig. \ref{Fig-1}B). On the other hand, the backward transitions are possible with the probability of order $\mathcal{O}(x)$ when the initial states are $(1)$ or $(2)$ as it requires only one backward transition. From the initial state $(3)$, from which the backward transition time is longer, the probability of making two backward steps is significantly lower, $\mathcal{O}(x^2)$. In other words, for the forward transitions, all the slow and fast paths are explored, but for the backward transitions only the fast path are utilized. Since the transition times are averages over all pathways, this leads to slower overall forward transitions. It can be shown that for more general case of not very small $x$ ($x<1$) these arguments are still valid, leading to the inequality $T_{\alpha}^{-}< T_{\alpha}^{+}$.

	Our observations on transition times has a very important consequence: it suggests that the asymmetry in the forward/backward transition rates can be used to quantitatively ``measure" crowding in molecular systems under non-equilibrium conditions. Crucially, this is a local measure that might provide a more comprehensive view on the microscopic mechanisms of underlying processes.

	While the non-equilibrium is a necessary condition for observing the asymmetry in the forward/backward transition times, it can be argued that this is not a sufficient condition. As has been shown above, the system with a single particle that prefers to move along the ring in the clockwise direction ($N=1$ system with arbitrary $L$ and $u>w$) is out of equilibrium, but the forward and backward transition times are the same: see Eq. (\ref{Eq-2}). To break the symmetry, the presence and interactions with other particles is required, suggesting that this phenomenon is driven by crowding in the system. It has been shown recently that the dynamics of the tracer particle should be non-Markovian for symmetry breaking in transition times to appear \cite{shin2020biased, shin2020asymmetry}. In our system, the non-Markovianity arises from the particle-particle interactions which is a consequence of crowding. The forward/backward symmetry breaking of the transition times discussed above should not be confused with the forward/backward symmetry breaking of the transition-path times discussed in Ref.\cite{berezhkovskii2019forward}.

\begin{figure}
		\centering
		\includegraphics[width=0.95\columnwidth]{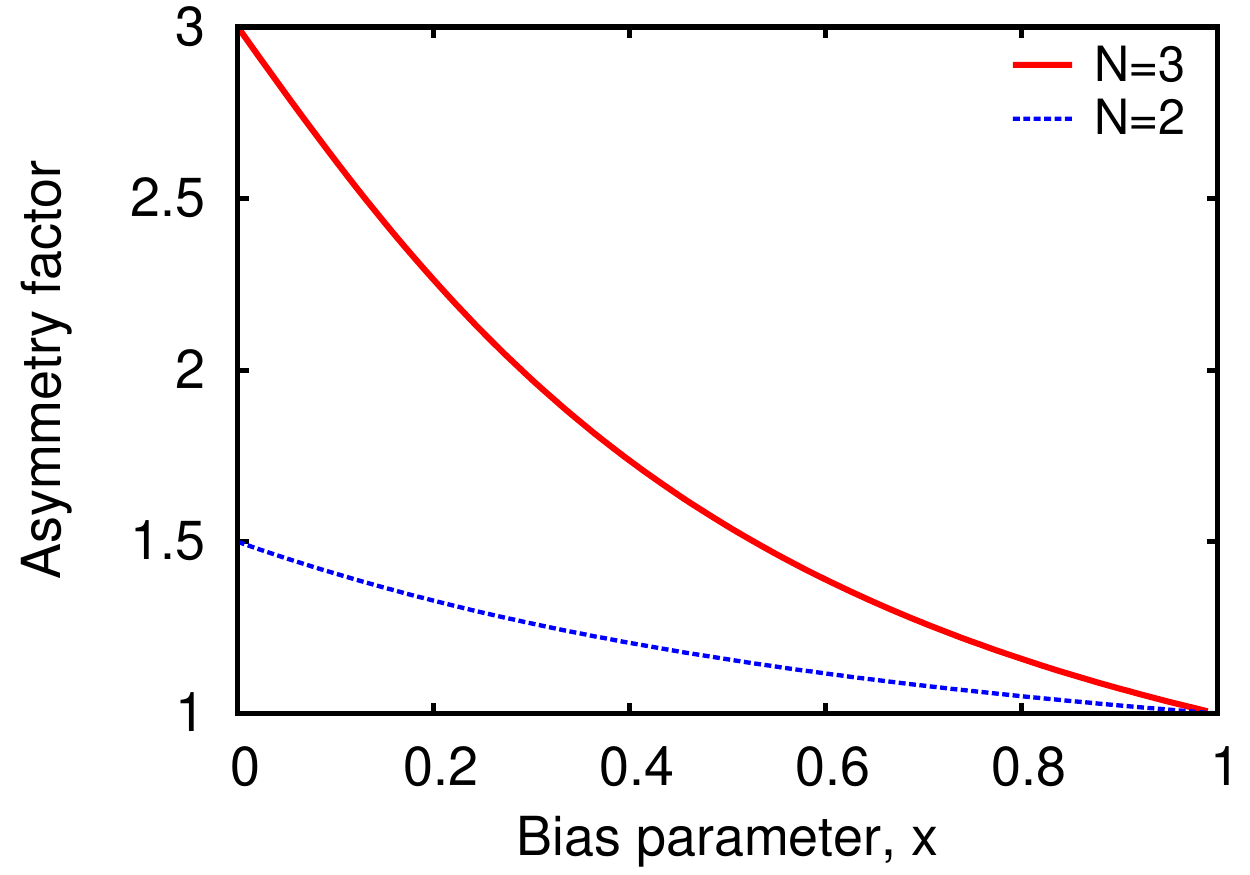}
		\caption{Asymmetry parameters as a function of the bias $x$. for two cases $L=4$, $N=3$ (Red solid line) and $L=4$, $N=2$ (Blue dashed line). Here we consider the time-averaging case. The explicit formulas for the case of $N=3$ are given in Ref. \cite{shin2020biased}.}
		\label{Fig-4}
	\end{figure}

	The above arguments suggest that the more frequent are interactions between the particles (larger the degree of crowding), the stronger is the deviation from the symmetry in the forward/backward transition times. Then we expect that varying the particle density (degree of crowding) in the system would modify the amplitude of symmetry breaking. In Fig. \ref{Fig-4} we show the asymmetry factor $A_{t}$ as a function of the bias parameter $x$ for two different particle densities ($N=2$ and $N=3$) on the lattice with $L=4$ sites for the time-averaging method of measurements.  For $N=3$ and $L=4$ the asymmetry factor has been already explicitly evaluated before \cite{shin2020biased}. As explained above, for $N=1$ we have $A_{\alpha}=1$. The results in Fig. \ref{Fig-4} show that the asymmetry magnitude $A_{t}$ increases with $N$ if other  parameters in the system are fixed. This finding highlights the observation that the asymmetry of the forward/backward transition times is originated from the particle-particle interactions, which is a signature of crowding in the system. Thus, measuring the asymmetry of forward/backward transition times in real molecular system would simultaneously provide an important microscopic information on the degree of crowding and on the deviation from the equilibrium. We speculate that measuring the transition dynamics of tracer particles at different locations might provide a comprehensive quantitative overall map of the system from the point of view of deviations from equilibrium and the strength of effective inter-particle interactions.
	
	\section{Summary and Conclusions}
	\label{Sec-summary}

	A theoretical investigation on the microscopic origin of symmetry breaking in dynamic properties of single particles in crowded systems is presented. By describing explicitly the dynamics of the tracer particle in one-dimensional discrete-state stochastic model, we evaluated transition probabilities, mean transition and residence times, and drift velocities. This allowed us to shed some light on the molecular mechanisms of asymmetry in the transition times along and against the bias. The theoretical analysis concludes that the symmetry breaking is driven by multi-particle interactions that appear due to crowding and by the degree of the system deviation from the equilibrium. It is also found that explicit values of all dynamic properties depend on how they are measured, and two specific methods, resetting and time-average, are identified. However, for both approaches the forward transition times (with respect to the local bias) are always longer than the backward transition times. All these observations are explained by explicitly accounting for different pathways for particle transitions in the stochastic model. 
	In equilibrium conditions, for each forward pathway there will be an identical backward pathway due to microscopic reversibility. On the contrary, in out-of-equilibrium conditions, a net current on the transition network can break this identity, yielding the asymmetry of forward/backward transition time. Furthermore, it is argued that measuring the asymmetry of transition times, as well as other dynamic properties, in real systems will simultaneously quantify the local deviations from the equilibrium and the degree of crowding. This might significantly improve the analysis of the microscopic mechanisms of complex natural processes.

	While we considered a random walk model on a one-dimensional periodic lattice with simple exclusion interactions, real systems can be much more complex in multiple ways, therefore it would be important to investigate more general cases. The list of potential future directions includes  i) studying the transition dynamics in higher dimensional systems; ii) considering more general particle-particle interactions in addition to simple exclusion interactions, and iii) analysing other dynamic properties such as cycle completion times as was done in Ref  \cite{vorac2020cycle}. In those cases, we expect that as long as the system is out of equilibrium and the tracer dynamics are non-Markovian (the crowding is present), the asymmetry of forward/ backward transition time would remain. But it will be important to quantify the degree of symmetry breaking. Lastly, we note that the symmetry of forward/backward transition times can also be broken in the presence of externally driven fluctuating forces as demonstrated experimentally in Ref. \cite{gladrow2019experimental}. Theoretical understanding of these phenomena would be an important direction of future studies.

	\section*{Acknowledgements}
	A.B.K. acknowledges the support from the Welch Foundation (C-1559), from the NSF (CHE-1953453 and MCB-1941106), and the Center for Theoretical Biological Physics sponsored by the NSF (PHY-2019745). A.M.B. was supported by the Intramural Research Program of the NIH, Center for Information Technology.
	
	\section*{Data Availability Statement}
	The data that support the findings of this study are available from the corresponding author upon reasonable request.

\end{document}